# Thermal conductance at the graphene-SiO$_2$ interface measured by optical pump-probe spectroscopy

Kin Fai Mak, Chun Hung Lui, and Tony F. Heinz[*]

*Departments of Physics and Electrical Engineering, Columbia University, 538 West 120$^{th}$ St., New York, NY 10027, USA*

## Abstract

We have examined the interfacial thermal conductance $\sigma_{int}$ of single and multi-layer graphene samples prepared on fused SiO$_2$ substrates by mechanical exfoliation of graphite. By using an ultrafast optical pump pulse and monitoring the transient reflectivity on the picosecond time scale, we obtained an average $\sigma_{int}$ of 5,000 W/cm$^2$K for the graphene-SiO$_2$ system. We observed significant variation in $\sigma_{int}$ between individual samples, but found no systematic dependence on the thickness of the graphene layers.

*Corresponding author: tony.heinz@columbia.edu

Graphene, a monolayer–thick sheet of graphite, has attracted much attention because of its unique electronic properties[1,2]. While our understanding of the electronic states and transport in graphene has advanced dramatically[1,2], our knowledge of thermal transport in this material system is currently less advanced[2-4]. The thermal transport properties are, however, of interest for both fundamental reasons and for applications. From the fundamental perspective, thermal transport provides insight into the properties of phonons and their interactions[5-7]. For applications, power dissipation often limits device performance[8]. Of particular importance is the issue of interfacial thermal transport between graphene layers and the substrate[9], which plays a critical role in current saturation of graphene devices[10,11].

In this letter, we report the determination of the interfacial thermal conductance $\sigma_{int}$ of graphene on a SiO$_2$ substrate. The measurements were performed using sudden heating of the graphene layers, exfoliated on transparent SiO$_2$ substrates, by a femtosecond laser pulse. The subsequent heat flow across the graphene-substrate interface, which occurs on the time sale of tens of picoseconds, was determined by following the temperature evolution of the graphene sample with a time-delayed optical probe pulse. From analysis of these data, we deduce an average value of $\sigma_{int}$ = 5,000 W/cm$^2$K for graphene on SiO$_2$. A significant variation in $\sigma_{int}$ was observed for different individual samples, with values as high as 11,000 W/cm$^2$K having been observed. On the other hand, in measuring graphene samples with layer thicknesses from $N$ = 1-13 monolayers, we found no systematic variation of $\sigma_{int}$ with thickness.

A Ti-sapphire laser operating at 80 MHz repetition rate was used to deliver 800 nm pump pulses of 100 fs duration. Frequency-doubled probe pulses were generated by a β-barium borate (BBO) crystal. The pump and probe beams were focused onto the samples with a 40× objective to obtain spot sizes of ~ 5-10 μm. Pump-probe measurements were then performed by modulating the pump laser at 1 kHz and detecting the synchronous change in probe reflection with a lock-in amplifier. The pump fluence was varied between 100 and 500 μJ/cm$^2$, with the probe fluence kept below 10% of this value. The induced modulation of the probe beam for monolayer graphene samples was ~ 10$^{-6}$.

The graphene samples were prepared by mechanical exfoliation[2] of kish graphite deposited on transparent SiO$_2$ substrates (Chemglass Inc.) cleaned in methanol. The surface flatness was characterized by atomic-force microscopy (AFM). A roughness of 1 – 2 nm was observed over a typical 10 × 10 μm$^2$ area, similar to that found on thermally grown silicon dioxide on Si surfaces[12]. The exfoliated graphene samples studied were of homogeneous layer thickness and areas of several hundreds to thousands of μm$^2$. The sample thickness was determined by optical absorption measurements, which provides monolayer accuracy[13]. All experimental measurements were carried out under ~ 10 mTorr vacuum at room temperature.

Figure 1 displays the representative transient reflectivity data for mono-, 8- and 13-layer graphene samples over a 200 ps time window. The decay of these transients can be fit using a bi-exponential form. The fast component has a time constant $\tau_1$ ~ 2 ps; the slow component has a thickness-dependent time constant $\tau_2$ ~ 10-100 ps. No pump fluence dependence of the decay dynamics was seen over different pump fluences (Fig. 1). This indicates that our measurement is performed in the regime of linear perturbation. Figure 2 summarizes the measured values for time constant $\tau_2$ for the slow component of the response as a function of sample layer thickness *N*. Multiple data points for one thickness correspond to measurements of different graphene samples of the given thickness.

For the purpose of this investigation, we are not concerned with the fast relaxation component. As discussed in earlier publications, this picosecond response is associated with the equilibration of the electronic excitations and optical phonons with other phonons in the system[14, 15]. On the time scale of the slow relaxation component, full equilibrium between the different degrees of freedom in the graphene samples should be achieved, with essentially all of the thermal energy should reside in excitations of acoustic phonons. On this time scale, we can then relate the reflectivity transient to the temporal evolution of graphene temperature. Given the very slight observed change in the reflectivity, we can assume that the change in sample temperature is linear in the change in optical reflectivity. The 400-nm probe wavelength was chosen because of its enhanced sensitivity to temperature arising from a temperature-dependent shift of the optical transition energies near the M-point in the graphene Brillouin zone[14].

Once the different sub-systems of the sample have reached thermal equilibrium with one another, subsequent heat dissipation can in principle be achieved by in-plane heat flow within the graphene and out-of-plane heat transfer to the substrate. Given the micron size of the laser spot, lateral heat flow will not be significant on the relevant (subnanosecond) time scale, since $\tau_{lat} = d^2 D_{gr,P}^{-1}$; $25 \mu m^2 (2.5 cm^2/s)^{-1}$; $100 ns$, where $\tau_{lat}$, $d$, and $D_{gr,P}$ denote, respectively, the lateral heat diffusion time, the laser spot size, and the in-plane thermal diffusivity of the sample (approximated by that of graphite[16]). In addition, cooling by lateral heat flow is incompatible with the observed dependence of $\tau_2$ on the thickness of the graphene sample (Fig. 2). The observed decay in temperature thus arises from heat flow into the substrate.

The rate of heat flow into the substrate is characterized by the values of interfacial thermal conductance $\sigma_{int}$. The values of $\sigma_{int}$ can be extracted from $\tau_2$ by considering simultaneously vertical heat flow within the graphene sample and the SiO$_2$ substrate, which we treat as diffusive, and heat flow across the interface (at $z = 0$) described by interfacial conductance $\sigma_{int}$. The initial condition for the heat-flow problem is defined by the nature of the optical excitation. As the optical penetration length $\alpha^{-1} \approx 15 nm$ [13] is much larger than the thickness of the graphene samples, the initial temperature is essentially spatially homogeneous. The actual graphene temperature just after laser heating can then be determined directly from the absorbed laser fluence and its specific heat capacity (again approximated by that of graphite[17]). For the highest applied fluence of $F = 500$ µJ/cm$^2$, we obtain a temperature rise of about 150 K. The substrate is completely transparent and remains homogeneously at room temperature during laser excitation. Using this initial condition, the heat dissipation problem can be solved numerically.

We fit the experimental data using the thermal diffusivities of graphite[16] (out-of-plane diffusivity of 0.017 cm$^2$/s) and fused silica[18] (0.009 cm$^2$/s) and treating $\sigma_{int}$ as an unknown parameter. We find (Fig. 2) that the calculated decay time constants $\tau_2$ vary nearly linearly with N. Particularly for the lower values of $\sigma_{int}$, interfacial heat flow completely controls cooling of the graphene samples. In this *interface dominated regime*, $\tau_2$ varies linearly with N, since the total amount of heat that must be transported across the interface increases linearly with the amount of materials. On the other hand, for sufficiently high values of $\sigma_{int}$, greater substrate heating for thicker graphene samples leads to reduced heat flow across the interface. The relation between $\tau_2$ and $N$ then deviates from linearity, as can also be seen in Fig. 2. Finally, in the *substrate dominated regime* where bulk heat flow completely defines the cooling of the sample, we obtain the $\tau_2 \propto N^2$, as expected for diffusive heat transport. Based on our experimental data (Fig. 2), we see that we are in fact essentially in the regime of interface limited heat flow.

In Fig. 3(a) we present a histogram of the values of the interfacial thermal conductance $\sigma_{int}$ inferred for graphene samples of all layer thicknesses. From a Gaussian fit, we obtain an average $<\sigma_{int}> = 5,000$ W/cm$^2$K (compatible with that deduced in ref. [9]), with a standard deviation of 1300 W/cm$^2$K. Fig. 3(b) displays the individual values of $\sigma_{int}$ from our measurements as a function of layer thickness $N$. Within our experimental accuracy, no correlation is seen. Further studies would be necessary to discern any possible systematic dependence of $\sigma_{int}$ on layer thickness.

The values of $\sigma_{int}$ obtained for the graphene-SiO$_2$ interface are comparable to those measured for single walled carbon nanotubes in solution[19] and on SiO$_2$ substrates[20]. The values are also similar to those reported for different metal-insulator interfaces, which typically lie between 3,000 and 11,000 W/cm$^2$K [21, 22]. While the graphene-SiO$_2$ interface displays good thermal transport properties, the measured values for $\sigma_{int}$ vary between 2,000 W/cm$^2$K and 11,000 W/cm$^2$K. This large dispersion presumably reflects the relatively poorly defined nature of interface produced by the mechanical deposition process. Some of the samples may have better surface contact with the substrate than others, yielding accordingly better interfacial heat flow. The highest value of the interfacial conductance obtained in this measurement was $\sigma_{int}$ = 11,000 W/cm$^2$K. The intrinsic value for the graphene-SiO$_2$ interface could be even higher.

We thank Drs. Yang Wu, Daohua Song, and Hugen Yan for fruitful discussions. We acknowledge support from the Nanoscale Science and Engineering Initiative of the National Science Foundation and from the MURI of the Air Force Office of Scientific Research.

**Figure captions:**

Fig. 1: Transient reflectivity decay dynamics of mono-, 8- and 13-layer graphene samples, together with double exponential fits to the data. The decay dynamics for the 13-layer graphene sample is plotted for two different pump fluences. After rescaling, identical decay dynamics are observed.

Fig. 2: Decay time constants for interfacial heat flow as a function of graphene layer thickness. Dots: experimental data. Lines: numerical calculations for different values of the interfacial thermal conductance $\sigma_{int}$.

Fig. 3: (a) Histogram of the values of the interfacial thermal conductance $\sigma_{int}$ extracted from the data in Fig. 2(a). Green curve: Gaussian fit to data, yielding conductance of $\sigma_{int}$ = 5,000 ± 1,300 W/cm$^2$K. (b) Measured $\sigma_{int}$ as a function of graphene layer thickness. No correlation with thickness is seen. The black horizontal line represents the average value.

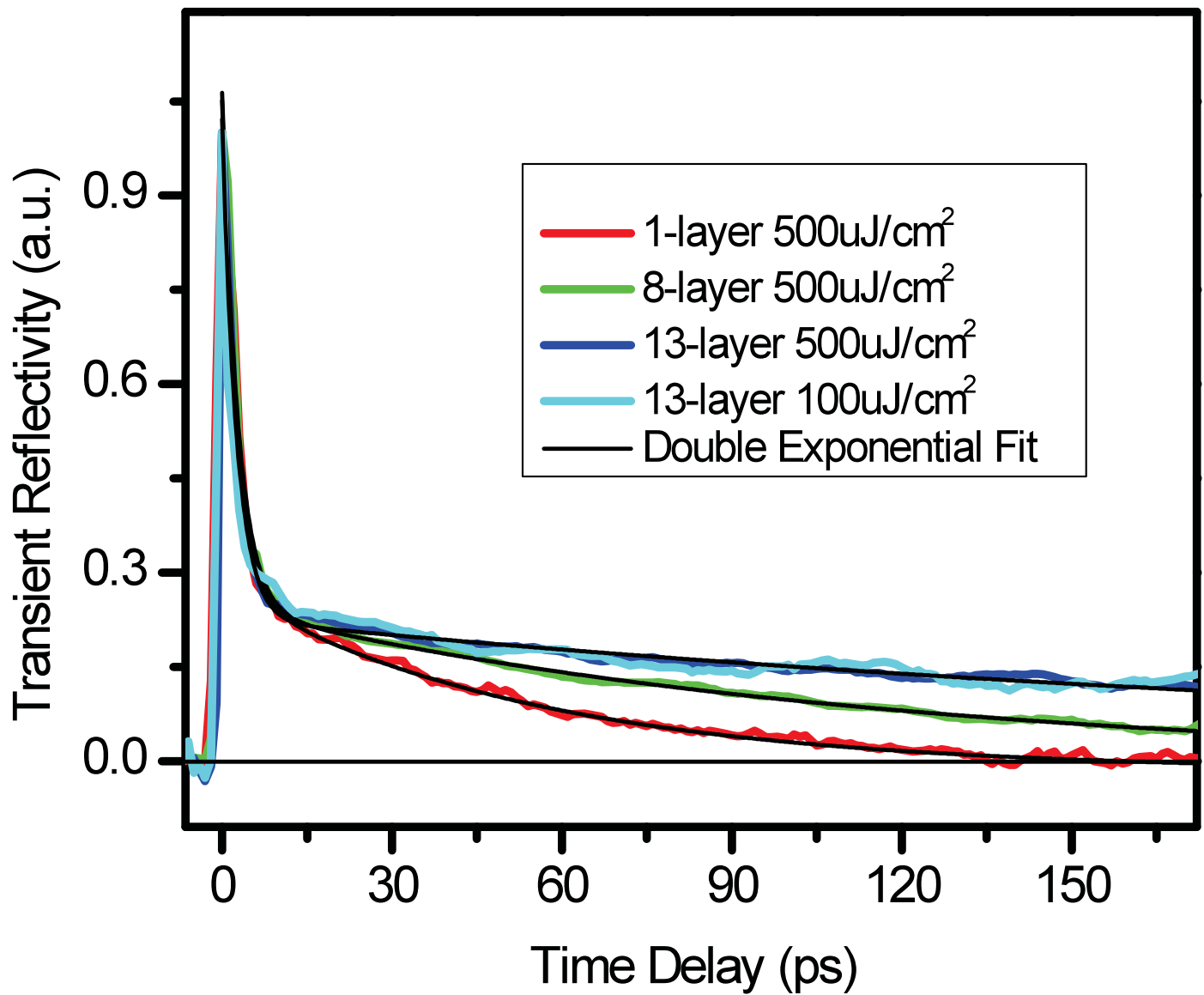

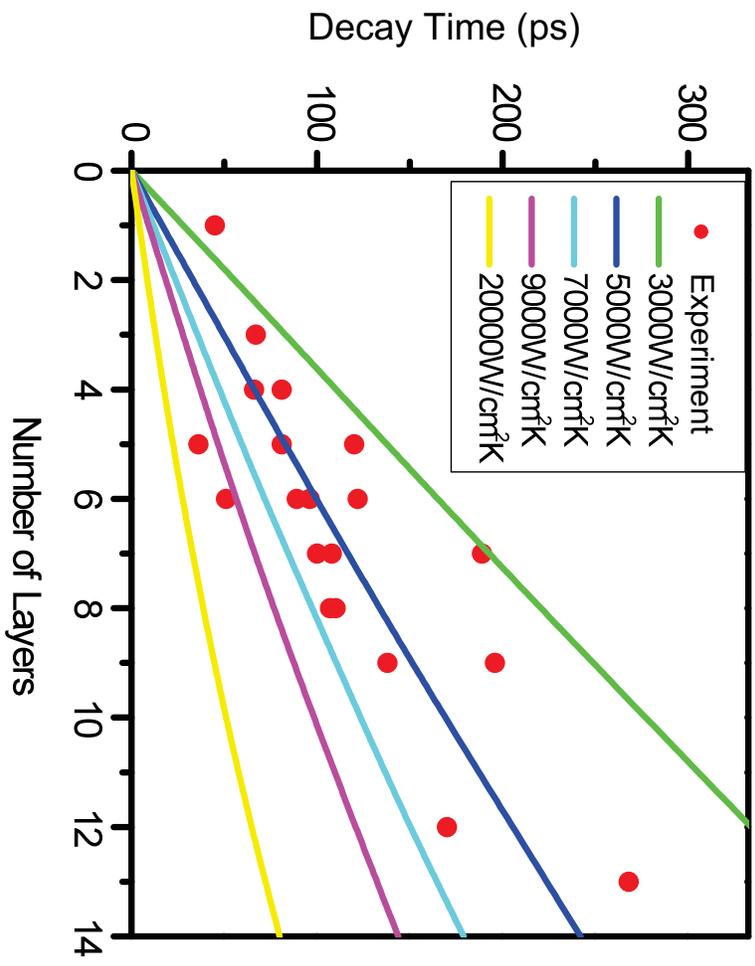

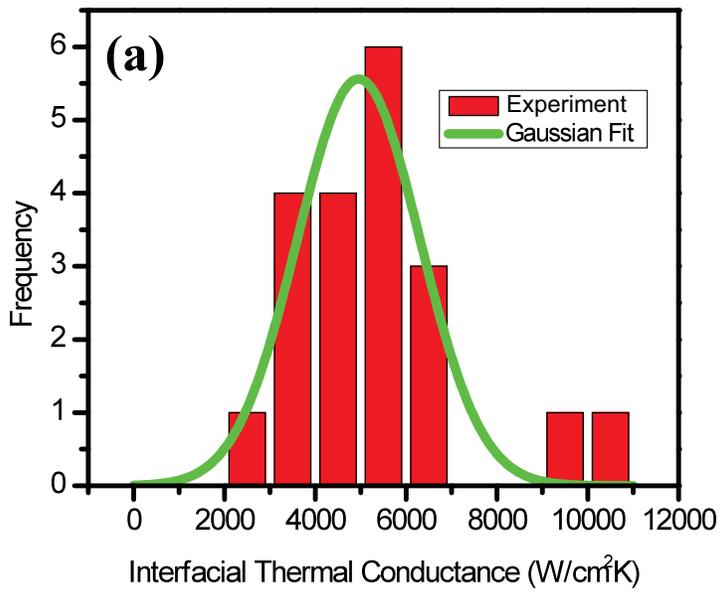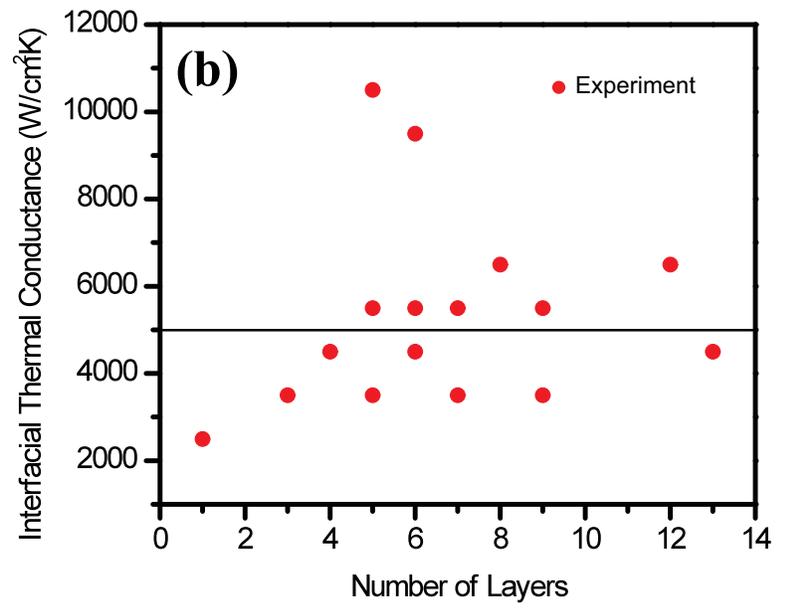